\documentclass[12pt]{article}
\newcommand{\dbox}{\rule {2mm}{2mm}}
\date{}

\begin{document}
\title{Relativistic Distance}
\author{ M. Akbari  AND
M. T. Darvishi$^*$\\
$^*$Dept. of Maths., Razi University, Kermanshah,
IRAN.\\darvishi@razi.ac.ir and darvishimt@yahoo.com }
 \maketitle
\begin{abstract}
 In this paper, we prove that two different observers don't
equally measure the distance between two points $A$ and $B$. For
this, we introduce some postulates and obtain a new formula to
show distance between $A$ and $B$. In this formula, radius of
universe, ${\bf n,}$ is entered such that if ${\bf n}$ tends to
infinity the ordinary distance is obtained.

\end{abstract}

\section {Introduction}

T. S. Eliot has said ''Each venture is a new beginning.'' We claim
this paper is a venture. This is a new theory about distance, if
radius of universe be finite. This theory can result in a
completely new view of the nature of distance. Ordinary
measurement  is so much a part of our daily life that almost
everyone has some conceptual difficulty in understanding our idea
of distance when he(she) first studies it.

Einstein may have put his finger on the difficulty when he said
\cite{resnick} ''Common sense is that layer of prejudices laid
down in the mind prior to the age of eighteen''. Indeed, it has
been said that every great theory begins as a heresy and ends as a
prejudice. More than a half-century of experimentation and
application has removed special relativity theory from the heresy
stage and put it on a sound conceptual and practical basis.

In this article, we show that there is a careful analysis of new
theory about distance. Furthermore, we obtain a formula to measure
distance between two points $x$ and $y$. In this study, we measure
distance between two points and compare it for two fixed
observers. For simplicity, we consider the problem in one
dimension. Assume that universe is bounded, however if the
universe is not bounded, universe radius has a finite value. Some
scientists such as Einstein and Rimman have believed unboundedness
of universe does not mean it is infinite \cite{einstein,einstein2}.

Einstein has said  ''Coming into existence of non-Euclidean
geometry resulted in this reality that infiniteness of space is
dubitable'' \cite{einstein}. He also has said ''Closed but
unbounded spaces are imaginable'' \cite{einstein}. We can measure
relative distance between two frames may be by comparing
measurement between frames. But then we have not deduced the
relative distance from observations confined to a single frame.

Furthermore, there is no way at all of determining the absolute
distance of an inertial reference frame. No inertial frame is
preferred over any other, for the laws of mechanics are the same
in all. We say that all inertial frame are equivalent as far as
mechanics is concerned.

\section {Postulates}

In this section, for our theory, we introduce some assumptions.
Their simplicity and generality are characteristic of this theory.
In this theory what is troublesome, apparently, is the philosophic
notion that length and time in the abstract are absolute
quantities and the belief that relativity contradicts this notion.
In this fitting, in emphasizing the common sense of relativity, to
conclude with this quotation from Bondi \cite{bondi} on the
presentation of relativity theory:\\
''At first, relativity was considered  shocking,
anti-establishment and highly mysterious, and all presentations
intended for the population at large were meant to emphasize these
shocking and mysterious aspects, which is hardly conductive to
easy teaching and good understanding.''

We explain our postulates as follows:
\begin{itemize}
  \item {\bf Postulate 1.} We can select any point as a reference frame
  (Relativity).
  \item {\bf Postulate 2.} Any point in itself frame lies in the center of
  universe  and measures ${\bf n}$ as the universe radius
  (Absolute infinity).
  \end{itemize}
By these postulates we introduce some principles as follows:
\begin{itemize}
  \item {\bf Principle 1.} Universe is symmetric in directions (Symmetry
  principle).
  \item {\bf Principle 2.} Every point agrees in the order of place of points
  (Agreement principle).
  \item {\bf Principle 3.} Consider three points $O_0,\,O_1$ and $A$ as
  the following order\\

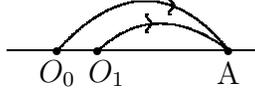
\unitlength=.6mm \special{em:linewidth 0.4pt}
\linethickness{0.4pt}
\begin{picture}(230.00,132.00)
\put(55.00,110.00){\rule{55 \unitlength}{.25\unitlength}}
\put(66.00,110.00){\circle*{1.50}}
\put(75.00,110.00){\circle*{1.50}}
\put(104.00,110.00){\circle*{1.50}}
\bezier{232}(66.00,110.00)(86.00,132.00)(104.00,110.00)
\bezier{148}(75.00,110.00)(88.00,122.00)(104.00,110.00)
\put(66.00,105.00){\makebox(0,0)[cc]{$O_0$}}
\put(77.00,105.00){\makebox(0,0)[cc]{$O_1$}}
\put(104.00,105.00){\makebox(0,0)[cc]{A}}
\bezier{20}(92.00,120.00)(90.00,122.00)(91.00,122.00)
\bezier{20}(92.00,120.00)(90.00,118.00)(91.00,118.00)
\bezier{20}(87.00,116.00)(85.00,118.00)(86.00,118.00)
\bezier{20}(87.00,116.00)(85.00,114.00)(86.00,114.00)
\end{picture}

\vspace{-5 cm}

\noindent
 then, distance of $O_0A$ in $O_0$ point of view is not
less than distance of $O_1A$ in $O_1$ point of view (Intuition
principle).
\end{itemize}
{\bf Notation.} Henceforth, if distance of $O_0A$ in $O_0$ point
of view is equal to $p$, we say that point view of $O_0$ from $A$
is $p$ and we will show  this by $(O_0\mapsto A)=p$. We show this
graphically by\\

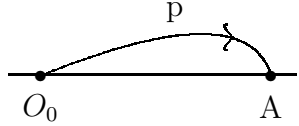
 \unitlength=.85mm \special{em:linewidth 0.4pt}
\linethickness{0.4pt}
\begin{picture}(101.00,125.00)
\put(50.00,112.00){\rule{45 \unitlength}{.25\unitlength}}
\put(55.00,112.00){\circle*{1.50}}
\put(91.00,112.00){\circle*{1.50}}
\put(91.00,107.00){\makebox(0,0)[cc]{A}}
\put(55.00,107.00){\makebox(0,0)[cc]{$O_0$}}
\bezier{192}(55.00,112.00)(87.00,125.00)(91.00,112.00)
\put(76.00,122.00){\makebox(0,0)[cc]{p}}
\bezier{16}(85.00,118.00)(83.00,120.00)(84.00,120.00)
\bezier{16}(85.00,118.00)(83.00,116.00)(84.00,116.00)
\end{picture}

\vspace{-8 cm}

 \noindent
{\bf Theorem.} If $N$ is a point such that $(O\mapsto N)={\bf n}$
for a fixed point $O$, then for any arbitrary point $O_1$ such
that $(O_1\mapsto N_1)={\bf n}$, $N_1$ in point view of $O_1$ is
$N$ in point view of $O$.\\
{\bf Proof.} Suppose that $O$ extends the universe as long as
${\bf n}$ to reach $N$. Also suppose that $O_1$ extends the
universe as long as ${\bf n}$ to reach $N_1$. If $N_1$ lies after
$N$ in point view of $O_1$ such as Figure $1$ \\

\unitlength=.6mm \special{em:linewidth 0.4pt}
\linethickness{0.4pt}
\begin{picture}(110.00,117.00)
\put(46.00,100.00){\circle*{1.50}}
\put(55.00,100.00){\circle*{1.50}}
\put(95.00,100.00){\circle*{1.50}}
\put(105.00,100.00){\circle*{1.50}}
\bezier{240}(55.00,100.00)(62.00,87.00)(105.00,100.00)
\bezier{244}(46.00,100.00)(55.00,117.00)(95.00,100.00)
\put(42.00,100.00){\rule{69 \unitlength}{.25\unitlength}}
\put(44.00,94.00){\makebox(0,0)[cc]{O}}
\put(55.00,94.00){\makebox(0,0)[cc]{$O_1$}}
\put(105.00,94.00){\makebox(0,0)[cc]{$N_1$}}
\put(99.00,103.00){\makebox(0,0)[cc]{N}}
\put(69.00,91.00){\makebox(0,0)[cc]{$\bf n$}}
\put(56.00,112.00){\makebox(0,0)[cc]{$\bf n$}}
\bezier{16}(72.00,108.00)(70.00,110.00)(71.00,110.00)
\bezier{16}(72.00,108.00)(70.00,106.00)(71.00,106.00)
\bezier{16}(83.00,95.00)(81.00,97.00)(82.00,97.00)
\bezier{16}(83.00,95.00)(81.00,93.00)(82.00,93.00)
\put(71.00,77.00){\makebox(0,0)[cc]{Figure 1}}
\end{picture}
\vspace{-4 cm}

\noindent
 by agreement principle $N_1$ in point view of $O$ lies
after $N$. Hence for $O$ the radius of universe is greater than
${\bf n}$, this contradicts with second postulate. Therefore,
$N_1$ in point view of $O_1$ lies before $N$ or coincides it. If
$N_1$ in point view of $O_1$ lies before $N$ then $O_1$ measures
the universe radius greater than ${\bf n}$. This contradicts with
second postulate. Consequently, $N_1$ in point view of $O_1$ is
$N$ in point view of $O$. \hspace {4cm} $\dbox$

\noindent
 {\bf Corollary.} For all points, $N$ is a fixed point.

 Consider the following figure, the fundamental question is: what
 is $(A \mapsto B)?$\\
\unitlength=.6mm \special{em:linewidth 0.4pt}
\linethickness{0.4pt}
\begin{picture}(120.00,130.00)
\bezier{5500}(40.00,100.00)(195.00,100.00)(40.00,100.00)
\put(45.00,100.00){\circle*{1.50}}
\put(55.00,100.00){\circle*{1.50}}
\put(75.00,100.00){\circle*{1.50}}
\put(85.00,100.00){\circle*{1.50}}
\put(115.00,100.00){\circle*{1.50}}
\put(44.00,96.00){\makebox(0,0)[cc]{$O_0$}}
\put(54.00,96.00){\makebox(0,0)[cc]{$O_1$}}
\put(75.00,96.00){\makebox(0,0)[cc]{A}}
\put(85.00,96.00){\makebox(0,0)[cc]{B}}
\put(115.00,96.00){\makebox(0,0)[cc]{N}}
\bezier{404}(45.00,100.00)(108.00,130.00)(115.00,100.00)
\bezier{284}(55.00,100.00)(103.00,116.00)(115.00,100.00)
\bezier{20}(102.00,114.00)(100.00,116.00)(101.00,116.00)
\bezier{16}(102.00,114.00)(100.00,112.00)(101.00,112.00)
\bezier{16}(98.00,108.00)(96.00,110.00)(97.00,110.00)
\bezier{20}(98.00,108.00)(96.00,106.00)(97.00,105.00)
\put(70.00,115.00){\makebox(0,0)[cc]{$\bf n$}}
\put(86.00,110.00){\makebox(0,0)[cc]{$\bf n$}}
\put(76.00,83.00){\makebox(0,0)[cc]{Figure 2}}
\end{picture}

\vspace{-4 cm}

\noindent
 In fact, we look for a formula to show distance of $A$ and $B$.
 Suppose
 $$(O_0 \mapsto O_1)=p, \ (O_0 \mapsto A)= x$$
 if $(O_1 \mapsto A) = x_1,$ we set $x_1= f(x,p)$ and we obtain function
 $f$ (Figure 3).\\

\unitlength=.60mm \special{em:linewidth 0.4pt}
\linethickness{0.4pt}
\begin{picture}(110.00,125.00)
\bezier{5500}(45.00,110.00)(190.00,110.00)(45.00,110.00)
\put(55.00,110.00){\circle*{1.50}}
\put(80.00,110.00){\circle*{1.50}}
\put(105.00,110.00){\circle*{1.50}}
\put(52.00,105.00){\makebox(0,0)[cc]{$O_0$}}
\put(82.00,106.00){\makebox(0,0)[cc]{$ O_1$}}
\put(108.00,105.00){\makebox(0,0)[cc]{A}}
\bezier{176}(55.00,110.00)(80.00,125.00)(80.00,110.00)
\bezier{172}(81.00,110.00)(105.00,125.00)(105.00,110.00)
\bezier{268}(55.00,110.00)(100.00,92.00)(105.00,110.00)
\put(71.00,99.00){\makebox(0,0)[cc]{$x$}}
\put(60.00,118.00){\makebox(0,0)[cc]{$p$}}
\put(117.00,120.00){\makebox(0,0)[cc]{$f(x,p)$}}
\bezier{16}(102.00,105.00)(99.00,104.00)(99.00,105.00)
\bezier{16}(102.00,105.00)(101.00,102.00)(102.00,102.00)
\bezier{20}(102.00,117.00)(100.00,120.00)(101.00,120.00)
\bezier{20}(102.00,117.00)(100.00,115.00)(101.00,115.00)
\bezier{20}(76.00,117.00)(74.00,120.00)(75.00,120.00)
\bezier{16}(76.00,117.00)(74.00,115.00)(75.00,115.00)
\put(83.00,85.00){\makebox(0,0)[cc]{Figure 3}}
\end{picture}

\vspace{-4 cm}

\noindent
 By above notations, in Figure 3 we have the following relations
 $$\left. \begin{array}{l}
   (O_0 \mapsto O_1)=p \\
   (O_0 \mapsto A)\ = x \\
 \end{array} \right \}\Rightarrow (O_1 \mapsto A)=f(p,x).
 $$
 If $(O_1 \mapsto O_0)=p'$, we have\\
\vspace{2.3cm}
 \unitlength=.60mm \special{em:linewidth 0.4pt}
\linethickness{0.4pt}
\begin{picture}(101.00,125.00)
\bezier{5500}(47.00,112.00)(150.00,112.00)(47.00,112.00)
\put(55.00,112.00){\circle*{1.50}}
\put(91.00,112.00){\circle*{1.50}}
\put(91.00,106.00){\makebox(0,0)[cc]{$O_1$}}
\put(52.00,106.00){\makebox(0,0)[cc]{$O_0$}}
\bezier{192}(55.00,112.00)(87.00,125.00)(91.00,112.00)
\put(76.00,122.00){\makebox(0,0)[cc]{$p$}}
\bezier{212}(55.00,112.00)(61.00,95.00)(91.00,112.00)
\put(73.00,99.00){\makebox(0,0)[cc]{$p'$}}
\bezier{16}(84.00,118.00)(82.00,120.00)(83.00,120.00)
\bezier{16}(84.00,118.00)(82.00,116.00)(83.00,116.00)
\bezier{20}(76.00,105.00)(78.00,108.00)(77.00,108.00)
\bezier{16}(76.00,105.00)(80.00,104.00)(79.00,104.00)
\put(75.00,85.00){\makebox(0,0)[cc]{Figure 4}}
\end{picture}

\vspace{-5 cm}

$$\left. \begin{array}{l}
   (O_0 \mapsto O_1)=p \\
   (O_0 \mapsto O_0) = 0 \\
 \end{array} \right \}\Rightarrow (O_1 \mapsto O_0)=f(p,0)
 $$
 hence $p'=f(p,0)$. Consider points $O_0$, $O_1$, $A$, $O'_1$ and
$A'$ such as Figure 5\\

\unitlength=.60mm \special{em:linewidth 0.4pt}
\linethickness{0.4pt}
\begin{picture}(130.00,113.00)
\bezier{5500}(40.00,100)(225.00,100)(40.00,100)
\put(45.00,100.00){\circle*{1.50}}
\put(65.00,100.00){\circle*{1.50}}
\put(85.00,100.00){\circle*{1.50}}
\put(105.00,100.00){\circle*{1.50}}
\put(125.00,100.00){\circle*{1.50}}
\put(128.00,95.00){\makebox(0,0)[cc]{$A$}}
\put(106.00,96.00){\makebox(0,0)[cc]{$O_1$}}
\put(86.00,95.00){\makebox(0,0)[cc]{$O_0$}}
\put(65.00,95.00){\makebox(0,0)[cc]{$O'_1$}}
\put(42.00,95.00){\makebox(0,0)[cc]{$A'$}}
\bezier{132}(45.00,100.00)(49.00,112.00)(65.00,100.00)
\bezier{232}(45.00,100.00)(53.00,81.00)(85.00,100.00)
\bezier{140}(65.00,100.00)(85.00,112.00)(85.00,100.00)
\bezier{136}(86.00,100.00)(105.00,112.00)(105.00,100.00)
\bezier{148}(105.00,100.00)(125.00,113.00)(125.00,100.00)
\bezier{248}(85.00,100.00)(125.00,82.00)(125.00,100.00)
\put(102.00,89.00){\makebox(0,0)[cc]{$x$}}
\put(49.00,89.00){\makebox(0,0)[cc]{$-x$}}
\put(39.00,111.00){\makebox(0,0)[cc]{ $f(-p,-x)$}}
\put(76.00,109.00){\makebox(0,0)[cc]{$-p$}}
\put(95.00,109.00){\makebox(0,0)[cc]{$p$}}
\put(136.00,107.00){\makebox(0,0)[cc]{$f(p,x)$}}
\bezier{16}(59.00,91.00)(61.00,93.00)(60.00,93.00)
\bezier{16}(59.00,91.00)(61.00,89.00)(60.00,89.00)
\bezier{16}(118.00,91.00)(116.00,93.00)(117.00,93.00)
\bezier{16}(118.00,91.00)(116.00,89.00)(117.00,89.00)
\bezier{16}(123.00,106.00)(121.00,108.00)(122.00,108.00)
\bezier{16}(123.00,106.00)(121.00,104.00)(122.00,104.00)
\bezier{16}(101.00,106.00)(99.00,108.00)(100.00,108.00)
\bezier{16}(101.00,106.00)(99.00,104.00)(100.00,104.00)
\bezier{16}(73.00,104.00)(75.00,106.00)(74.00,106.00)
\bezier{16}(73.00,104.00)(75.00,103.00)(74.00,103.00)
\bezier{16}(50.00,106.00)(52.00,108.00)(51.00,108.00)
\bezier{16}(50.00,106.00)(52.00,104.00)(51.00,104.00)
\put(83.00,78.00){\makebox(0,0)[cc]{Figure 5}}
\end{picture}

\vspace{-4 cm}

\noindent
 $$\left. \begin{array}{l}
   (O_0 \mapsto O_1)=p \\
   (O_0 \mapsto A)\ = x \\
 \end{array} \right \}\Rightarrow (O_1 \mapsto A)=f(p,x)
 $$

  $$\left. \begin{array}{l}
   (O_0 \mapsto O'_1)=-p \\
   (O_0 \mapsto A')\ = -x \\
 \end{array} \right \}\Rightarrow (O'_1 \mapsto A')=f(-p,-x).
 $$
By symmetric principle, universe is symmetric in directions. This
results in
$$(O_1 \mapsto A)\ = -(O'_1 \mapsto A')$$
or
\begin{equation}\label{no1}
  f(p,x)=-f(-p,-x).
\end{equation}
Hence function $f$ must have property (\ref{no1}).
 Consider $ (O_0 \mapsto O_1)=p$ and $ (O_1
\mapsto O_0)=p'.$ If $|p|=|p'|$ there isn't anything to prove.
Suppose that $|p|<|p'|,$ hence there is a point, for example,
$O'$, such that $ (O_1 \mapsto O')=-p.$ As  $|p|<|p'|,$ in $O_1$
point of view $O_0$ lies between $O_1$ and $O'$. Therefore by
agreement principle, for any other points $O_0$ lies between $O_1$
and $O'$ (Figure 6). Now, we obtain $ (O' \mapsto O_1)$\\

\unitlength=.60mm \special{em:linewidth 0.4pt}
\linethickness{0.4pt}
\begin{picture}(110.00,124.00)
\bezier{5500}(38.00,100)(185.00,100)(38.00,100)
\put(45.00,100.00){\circle*{1.50}}
\put(60.00,100.00){\circle*{1.50}}
\put(100.00,100.00){\circle*{1.50}}
\put(103.00,95.00){\makebox(0,0)[cc]{$O_1$}}
\put(58.00,96.00){\makebox(0,0)[cc]{$O_0$}}
\put(41.00,95.00){\makebox(0,0)[cc]{$O'$}}
\bezier{464}(45.00,100.00)(107.00,124.00)(100.00,100.00)
\bezier{432}(45.00,100.00)(43.00,78.00)(100.00,100.00)
\bezier{308}(60.00,100.00)(104.00,111.00)(100.00,100.00)
\bezier{372}(60.00,100.00)(67.00,91.00)(100.00,100.00)
\put(60.00,112.00){\makebox(0,0)[cc]{$-p'$}}
\put(75.00,106.00){\makebox(0,0)[cc]{$p$}}
\put(66.00,93.00){\makebox(0,0)[cc]{$p'$}}
\put(79.00,89.00){\makebox(0,0)[cc]{$-p$}}
\bezier{16}(52.00,90.00)(54.00,92.00)(53.00,92.00)
\bezier{16}(52.00,90.00)(54.00,88.00)(53.00,88.00)
\bezier{20}(52.00,90.00)(54.00,88.00)(53.00,88.00)
\bezier{16}(73.00,96.00)(75.00,98.00)(74.00,98.00)
\bezier{16}(73.00,95.00)(75.00,93.00)(74.00,93.00)
\bezier{16}(90.00,105.00)(88.00,107.00)(89.00,107.00)
\bezier{16}(89.00,105.00)(87.00,103.00)(88.00,103.00)
\bezier{16}(88.00,112.00)(86.00,114.00)(87.00,114.00)
\bezier{16}(88.00,112.00)(86.00,110.00)(87.00,110.00)
\put(66.00,77.00){\makebox(0,0)[cc]{Figure 6}}
\end{picture}

\vspace{-4 cm}

\noindent
  $$\left. \begin{array}{l}
   (O_1 \mapsto O')=-p \\
   (O_1 \mapsto O_1) =0 \\
 \end{array} \right \}\Rightarrow (O' \mapsto O_1)=f(-p,0)
 $$
 as $f$ has property (\ref{no1}) we have

  $$\left. \begin{array}{l}
   (O' \mapsto O_1)=-f(-(-p),0) \\
   (O' \mapsto O_1)\ = -f(p,0) \\
 \end{array} \right .
 $$
 or $p'=f(p,0)$ therefore
 $$ (O' \mapsto O_1)=-p'. $$
As in Figure 6 order of $O_1,\ O_0$ and $O'$ are similar for any
other points, and so as $|p|<|p'|,$ we have
 $$\left. \begin{array}{l}
   (O' \mapsto O_1)=-p' \\
   (O_0 \mapsto O_1) = p \\
   |p|\ < \ |p'|\\
 \end{array} \right \}\Rightarrow |(O' \mapsto O_1)| <  |(O_0 \mapsto
 O_1)|
 $$
 this contradicts the intuition principle. Therefore  the case of $ |p| <
 |p'|$ is not correct. Similarly assumption of $ |p'| <  |p|$
 results in a contradiction. Hence, $ p' = -p,$ or
\begin{equation}\label{no2}
  f(p,0)=-p.
\end{equation}
Consequently, function $f$ must have property (\ref{no2}).
Therefore we have
$$ (A\mapsto B)=-(B \mapsto A).$$

Now we obtain an important condition by Postulate $1$. Consider
$O_0, \ O_1,\ O_2$ and $A$ as the following figure \\

\unitlength=.70mm \special{em:linewidth 0.4pt}
\linethickness{0.4pt}
\begin{picture}(113.00,104.00)
\bezier{5500}(40.00,90)(200.00,90)(40.00,90)
\put(45.00,90.00){\circle*{1.50}}
\put(70.00,90.00){\circle*{1.50}}
\put(90.00,90.00){\circle*{1.50}}
\put(109.00,90.00){\circle*{1.50}}
\put(112.00,86.00){\makebox(0,0)[cc]{$A$}}
\put(87.00,86.00){\makebox(0,0)[cc]{$O_2$}}
\put(67.00,86.00){\makebox(0,0)[cc]{$O_1$}}
\put(43.00,86.00){\makebox(0,0)[cc]{$O_0$}}
\bezier{164}(45.00,90.00)(71.00,102.00)(70.00,90.00)
\bezier{140}(70.00,90.00)(93.00,104.00)(90.00,90.00)
\bezier{152}(90.00,90.00)(110.00,104.00)(109.00,90.00)
\bezier{200}(70.00,90.00)(86.00,76.00)(109.00,90.00)
\bezier{384}(45.00,90.00)(110.00,64.00)(109.00,90.00)
\put(65.00,78.00){\makebox(0,0)[cc]{$x_0$}}
\put(51.00,97.00){\makebox(0,0)[cc]{$p_0$}}
\put(77.00,98.00){\makebox(0,0)[cc]{$p_1$}}
\put(97.00,98.00){\makebox(0,0)[cc]{$x_2$}}
\put(103.00,83.00){\makebox(0,0)[cc]{$x_1$}}
\bezier{16}(105.00,97.00)(103.00,99.00)(104.00,99.00)
\bezier{16}(105.00,97.00)(103.00,95.00)(104.00,95.00)
\bezier{16}(87.00,97.00)(85.00,99.00)(86.00,99.00)
\bezier{16}(86.00,97.00)(84.00,95.00)(85.00,95.00)
\bezier{16}(66.00,96.00)(64.00,98.00)(65.00,98.00)
\bezier{16}(65.00,96.00)(63.00,94.00)(64.00,94.00)
\bezier{16}(94.00,84.00)(92.00,86.00)(93.00,86.00)
\bezier{16}(94.00,84.00)(92.00,82.00)(93.00,82.00)
\bezier{16}(101.00,78.00)(99.00,80.00)(100.00,80.00)
\bezier{16}(101.00,78.00)(99.00,76.00)(100.00,76.00)
\put(73.00,65.00){\makebox(0,0)[cc]{Figure 7}}
\end{picture}

\vspace{-4 cm}

\noindent
 we select $O_0$ as the reference frame, so
\begin{equation} \label {no3} \left.
\begin{array}{l}
   (O_0 \mapsto O_1)=p_0 \\
   (O_0 \mapsto A) =x_0 \\
 \end{array} \right \}\Rightarrow (O_1 \mapsto A)=f(p_0,x_0)=x_1.
 \end {equation}

\noindent
 We can select $O_1$ as the reference frame, by postulate $1$ we
 have
\begin{equation} \label {no4} \left.
\begin{array}{l}
   (O_1 \mapsto O_2)=p_1 \\
   (O_1 \mapsto A) =x_1 \\
 \end{array} \right \}\Rightarrow (O_2 \mapsto A)=f(p_1,x_1)=x_2
 \end {equation}

\noindent
 from (\ref{no3}) and (\ref{no4}) we have
\begin{equation}\label{no5}
  f(p_1,f(p_0,x_0))=x_2=(O_2 \mapsto A)
\end{equation}
also
\begin{equation}\label{no6}
  (O_0 \mapsto O_1)= p_0 \Rightarrow \left . \begin{array}{l}
   (O_1 \mapsto O_0)=-p_0 \\
   (O_1 \mapsto O_2) =p_1 \\
 \end{array} \right \}\Rightarrow (O_0 \mapsto O_2)=f(-p_0,p_1)
\end{equation}
and so $(O_0 \mapsto A) =x_0$, therefore
\begin{equation}\label{no7} \left . \begin{array}{l}
  (O_0 \mapsto O_2)=f(-p_0,p_1) \\
   (O_0 \mapsto A) =x_0 \\
 \end{array} \right \}\Rightarrow (O_2 \mapsto
 A)=f(f(-p_0,p_1),x_0)=x_2.
\end{equation}
Finally, from (\ref{no5}) and (\ref{no7}) we have
\begin{equation}\label{no8}
f(f(-p_0,p_1),x_0)=f(p_1,f(p_0,x_0)).
\end{equation}
Thus function $f$ must have property (\ref{no8}). Consider the
following figure\\

\unitlength=.60mm \special{em:linewidth 0.4pt}
\linethickness{0.4pt}
\begin{picture}(121.00,113.00)
\bezier{5500}(40.00,100)(200.00,100)(40.00,100)
\put(45.00,100.00){\circle*{1.50}}
\put(60.00,100.00){\circle*{1.50}}
\put(115.00,100.00){\circle*{1.50}}
\put(119.00,96.00){\makebox(0,0)[cc]{$N$}}
\put(60.00,96.00){\makebox(0,0)[cc]{$O_1$}}
\put(41.00,96.00){\makebox(0,0)[cc]{$O_0$}}
\bezier{88}(45.00,100.00)(55.00,108.00)(60.00,100.00)
\bezier{276}(60.00,100.00)(115.00,113.00)(115.00,100.00)
\bezier{352}(45.00,100.00)(56.00,78.00)(115.00,100.00)
\put(47.00,105.00){\makebox(0,0)[cc]{$p$}}
\put(75.00,107.00){\makebox(0,0)[cc]{$\bf n$}}
\put(98.00,90.00){\makebox(0,0)[cc]{$\bf n$}}
\bezier{16}(110.00,106.00)(108.00,108.00)(109.00,108.00)
\bezier{16}(110.00,106.00)(108.00,104.00)(109.00,104.00)
\bezier{16}(85.00,91.00)(83.00,93.00)(84.00,93.00)
\bezier{16}(85.00,91.00)(83.00,89.00)(84.00,89.00)
\bezier{16}(55.00,104.00)(53.00,106.00)(54.00,106.00)
\bezier{16}(55.00,104.00)(53.00,102.00)(54.00,102.00)
\put(80.00,78.00){\makebox(0,0)[cc]{Figure 8}}
\end{picture}

\vspace{-4 cm}

\noindent
 by second postulate there are two points on an axis such
that all points lie between and middle them. We call right hand
side point as $N$. For any point such as $O$ we have $(O \mapsto
N)= {\bf n}$. Suppose that $(O_0 \mapsto O_1) =p$ and $O_0$ is the
reference frame. Therefore
\begin{equation} \label {no9} \left.
\begin{array}{l}
   (O_0 \mapsto O_1)=p \\
   (O_0 \mapsto N) ={\bf n} \\
 \end{array} \right \}\Rightarrow (O_1 \mapsto N)=f(p,{\bf n})={\bf n}.
 \end {equation}

 Hence function $f$ must have property (\ref{no9}).
We summarize conditions of $f$ as follows
\begin{equation}\label{no10}
  f(x,y)=-f(-x,-y).
\end{equation}
\begin{equation}\label{no11}
  f(x,0)=-x
\end{equation}
\begin{equation}\label{no12}
  f(f(x,y),z)=f(y,f(-x,z))
 \end{equation}
\begin{equation}\label{no13}
  f(x,{\bf n})={\bf n}.
\end{equation}

 \section {Finding Function $f$}

 In this section we present function $f(x,y)$ with properties
 (\ref{no10}) up to (\ref{no13}). We propose the following function
 which satisfies conditions (\ref{no10}) up to (\ref{no12})
\begin{equation}
  f(x,y)=\frac{y-x}{1+\beta xy}\cdot
\end{equation}
To obtain $\beta$ we use condition (\ref{no13}), that is
$$f(x,{\bf n})={\bf n}$$
thus
$$\frac{{\bf n}-x}{1+\beta x{\bf n}}={\bf n}$$
for this $\beta=- \frac 1{{\bf n}^2} \cdot$ Therefore the final
form of $f$ is
\begin{equation}\label{ffinal}
  f(x,y)=\frac{y-x}{1-\frac{xy}{{\bf n}^2}}\cdot
\end{equation}
{\bf Remark}. For $n \rightarrow \pm \infty$ relation
(\ref{ffinal}) is ordinary directed distance of $x$ and $y$.

\section {Examples}

In this section we present some examples. In these examples we
obtain relative distance of two particular points. \\
Example 1. Consider three points $O_0, \ O_1$ and $A$ as the
following figure\\

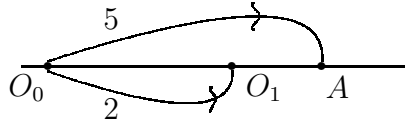
\unitlength=.70mm \special{em:linewidth 0.4pt}
\linethickness{0.4pt}
\begin{picture}(100.00,108.00)
\bezier{5500}(40.00,90)(185.00,90)(40.00,90)
\put(45.00,90.00){\circle*{1.50}}
\put(80.00,90.00){\circle*{1.50}}
\put(97.00,90.00){\circle*{1.50}}
\put(100.00,86.00){\makebox(0,0)[cc]{$A$}}
\put(86.00,86.00){\makebox(0,0)[cc]{$O_1$}}
\put(41.00,86.00){\makebox(0,0)[cc]{$O_0$}}
\bezier{196}(45.00,89.00)(80.00,77.00)(80.00,89.00)
\bezier{292}(45.00,91.00)(98.00,108.00)(97.00,91.00)
\put(57.00,99.00){\makebox(0,0)[cc]{$5$}}
\put(57.00,82.00){\makebox(0,0)[cc]{$2$}}
\bezier{16}(85.00,100.00)(83.00,102.00)(84.00,102.00)
\bezier{16}(85.00,99.00)(83.00,97.00)(84.00,97.00)
\bezier{16}(77.00,84.00)(75.00,86.00)(76.00,86.00)
\bezier{28}(77.00,84.00)(75.00,80.00)(76.00,82.00)
\end{picture}

\vspace{-5 cm}

\noindent
 what is $(O_1 \mapsto A)?$ From our notation we
have

 $$\left. \begin{array}{l}
   (O_0 \mapsto O_1)=2 \\
   (O_0 \mapsto A) =5 \\
 \end{array} \right \}\Rightarrow (O_1 \mapsto A)=f(2,5)
 $$
 thus
 $$(O_1 \mapsto A)=\frac{5-2}{1-\frac{10}{{\bf n}^2}}=\frac {3 {\bf
 n}^2}{{\bf n}^2-10} \cdot$$
 And so
 $$\left. \begin{array}{l}
   (O_1 \mapsto O_0)=-2 \\
   (O_1 \mapsto A) =\frac {3 {\bf
 n}^2}{{\bf n}^2-10}  \\
 \end{array} \right \}\Rightarrow (O_0 \mapsto A)=f(-2,\frac {3 {\bf
 n}^2}{{\bf n}^2-10})
 $$
 or
$$ (O_0 \mapsto A)=\frac
{\frac {3 {\bf n}^2}{{\bf n}^2-10}
 +2}
 {1+
 \frac{6{\bf n}^2}{{\bf n}^2({\bf
 n^2}-10)}}=\frac{5({\bf n}^2-4)}{{\bf n}^2-4}=5$$
 such that we expect.\\
 Example 2. Suppose that $N_1$ is a point near $N$, e.g., for
 $O_0$ we have $(O_0 \mapsto N_1)={\bf n}- \epsilon$, for any positive and small
 $\epsilon$. Does
 mentioned theorem in section 2 hold? In other words does $(N_1 \mapsto N)={\bf
 n}$? We have
$$  \left.
\begin{array}{l}
   (O_0 \mapsto N_1)= {\bf n}- \epsilon\\
   (O_0 \mapsto N) ={\bf n} \\
 \end{array} \right \}\Rightarrow (N_1 \mapsto N)=f({\bf n}- \epsilon,{\bf n})
$$
 hence
 $$(N_1 \mapsto N)= \frac {{\bf n}-{\bf n}+ \epsilon}{1-\frac{{\bf
 n}({\bf n}- \epsilon)}{{\bf n}^2}}={\bf n}$$
 that is a surprising result.\\
 Example 3. What is point view of $N$ from $N$? In this situation
 we have an ambiguity. On the one hand, the distance of $N$ from
 itself is zero and on the other hand this is ${\bf n}$. Now by
 our notations
$$  \left.
\begin{array}{l}
   (O_0 \mapsto N)= {\bf n}\\
   (O_0 \mapsto N) ={\bf n} \\
 \end{array} \right \}\Rightarrow (N \mapsto N)=f({\bf n},{\bf n})$$
 or
 $$(N \mapsto N)=\frac{{\bf n}-{\bf n}}{1-\frac{{\bf n}^2}{{\bf
 n}^2}}=\frac{0}{0}$$
 which we forecasted it by our ambiguity.\\
 In the last example we show distance for $N$ has no sense.\\
 Example 4. Consider three points $O_0, \ O_1$ and $N$ as
 follows\\

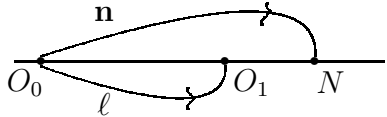
 \unitlength=.70mm \special{em:linewidth 0.4pt}
\linethickness{0.4pt}
\begin{picture}(100.00,108.00)
\bezier{5500}(40.00,90)(180.00,90)(40.00,90)
\put(45.00,90.00){\circle*{1.50}}
\put(80.00,90.00){\circle*{1.50}}
\put(97.00,90.00){\circle*{1.50}}
\put(100.00,86.00){\makebox(0,0)[cc]{$N$}}
\put(85.00,86.00){\makebox(0,0)[cc]{$O_1$}}
\put(42.00,86.00){\makebox(0,0)[cc]{$O_0$}}
\bezier{196}(45.00,89.00)(80.00,77.00)(80.00,89.00)
\bezier{292}(45.00,91.00)(98.00,108.00)(97.00,91.00)
\put(57.00,99.00){\makebox(0,0)[cc]{$ \bf n$}}
\put(57.00,82.00){\makebox(0,0)[cc]{$\ell$}}
\bezier{16}(89.00,99.00)(87.00,101.00)(88.00,101.00)
\bezier{16}(88.00,99.00)(86.00,97.00)(87.00,97.00)
\bezier{16}(74.00,83.00)(72.00,85.00)(73.00,85.00)
\bezier{16}(74.00,83.00)(72.00,81.00)(73.00,81.00)
\end{picture}

\vspace{-5 cm}

\noindent
 what are $(N \mapsto O_0)$ and $(N \mapsto O_1)$? we
have
 $$  \left.
\begin{array}{l}
   (O_0 \mapsto N)= {\bf n}\\
   (O_0 \mapsto O_0) =0 \\
 \end{array} \right \}\Rightarrow (N \mapsto O_0)=f({\bf n},0)$$
 hence
 $$(N \mapsto O_0)=\frac{0-{\bf n}}{1-0}=-{\bf n}$$
 and so
$$  \left.
\begin{array}{l}
   (O_0 \mapsto N)= {\bf n}\\
   (O_0 \mapsto O_1) =\ell \\
\end{array} \right \}\Rightarrow (N \mapsto O_1)=f({\bf n},\ell)$$
thus
$$(N \mapsto O_1)=\frac{\ell-{\bf n}}{1-\frac{{\bf n}\ell}{{\bf n}^2}}=-{\bf n}.$$
Therefore point view of $N$ from distance of $O_0$ and $O_1$
is the difference of $(N \mapsto O_0)$ and $(N \mapsto O_1)$, or
$$ -{\bf n}+{\bf n}=0.$$
Consequently distance for $N$ has no sense.

\end{document}